\documentclass[10pt,twocolumn,letterpaper]{article}

\usepackage{cvpr}
\usepackage{times}
\usepackage{epsfig}
\usepackage{graphicx}
\usepackage{amsmath}
\usepackage{amssymb}
\usepackage{multirow}

\usepackage[tight,footnotesize]{subfigure}
\usepackage[ruled,vlined]{algorithm2e}

\usepackage[pagebackref=true,breaklinks=true,letterpaper=true,colorlinks,bookmarks=false]{hyperref}

\cvprfinalcopy 

\ifcvprfinal\fi
\begin{document}

\title{StandardGAN: Multi-source Domain Adaptation for Semantic Segmentation of Very High Resolution Satellite Images by Data Standardization}

\author{Onur Tasar$^1$ ~ Yuliya Tarabalka$^2$ ~ Alain Giros$^3$ ~ Pierre Alliez$^1$ ~ S{\'e}bastien Clerc$^4$\\
$^1$Universit{\'e} C{\^o}te d'Azur, Inria ~ $^2$LuxCarta ~ $^3$Centre National d'{\'E}tudes Spatiales ~ $^4$ACRI-ST\\
{\tt\small onur.tasar@inria.fr}
}

\maketitle

\begin{abstract}
Domain adaptation for semantic segmentation has recently been actively studied to increase the generalization capabilities of deep learning models. The vast majority of the domain adaptation methods tackle single-source case, where the model trained on a single source domain is adapted to a target domain. However, these methods have limited practical real world applications, since usually one has multiple source domains with different data distributions. In this work, we deal with the multi-source domain adaptation problem. Our method, namely StandardGAN, standardizes each source and target domains so that all the data have similar data distributions. We then use the standardized source domains to train a classifier and segment the standardized target domain. We conduct extensive experiments on two remote sensing data sets, in which the first one consists of multiple cities from a single country, and the other one contains multiple cities from different countries. Our experimental results show that the standardized data generated by StandardGAN allow the classifiers to generate significantly better segmentation.

\end{abstract}

\section{Introduction}\label{sec:intro}

\begin{figure}
\centering
\subfigure[City A]{%
    \includegraphics[width=0.32\linewidth]{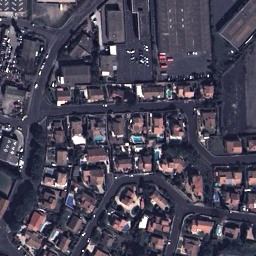}}
\subfigure[City B]{%
    \includegraphics[width=0.32\linewidth]{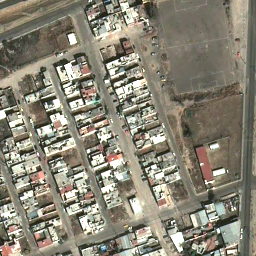}}
\subfigure[City C]{%
    \includegraphics[width=0.32\linewidth]{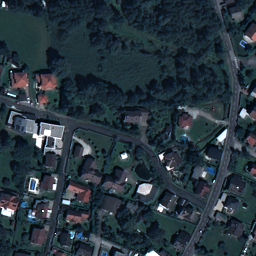}}
\subfigure[Standardized (a)]{%
    \includegraphics[width=0.32\linewidth]{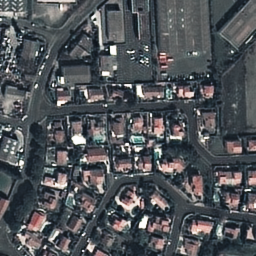}}
\subfigure[Standardized (b)]{%
    \includegraphics[width=0.32\linewidth]{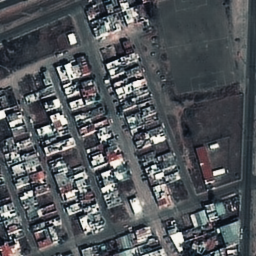}}
\subfigure[Standardized (c)]{%
    \includegraphics[width=0.32\linewidth]{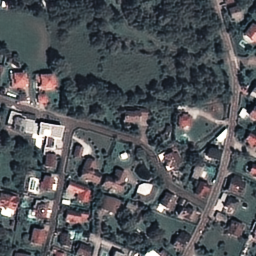}}
\caption{Real cities and the standardized data generated by StandardGAN.}
\label{fig:teaser}
\end{figure}

Over the years, semantic segmentation of remote sensing data has become an important research topic, due to its wide range of applications such as navigation, autonomous driving, and automatic mapping. In the last decade, a significant progress has been made, especially after \textit{convolutional neural networks (CNNs)} had revolutionized the computer vision community. Among CNNs, \textit{U-net}~\cite{ronneberger2015u} has gained an increasing attention due to its capability to generate highly precise semantic segmentation from remote sensing data. 

Nonetheless, it is a known issue that the performance of U-net or other CNNs immensely depends on the representativeness of the training data~\cite{tuia2016domain}. However, in remote sensing, having data that are representative to classify the whole world is challenging, because various atmospheric effects, intra-class variations, and differences in acquisition usually cause the images collected over different locations to have largely different data distributions. Such differences induce CNNs to generate unsatisfactory segmentation. This problem is referred to as \textit{domain adaptation} in the literature~\cite{tuia2016domain}. One way to overcome this issue is to manually annotate a small portion of test data to \textit{fine-tune} the already trained classifier~\cite{maggiori2016convolutional}. However, every time when new data are received, annotating even a small portion of them is labor-intensive.

Oftentimes, it is a good practice to perform \textit{data augmentation}~\cite{buslaev2018albumentations} to enlarge the training data and to reduce the risk of over-fitting. For example, in remote sensing, color jittering with random gamma correction or random contrast change is commonly used~\cite{tasar2019incremental}. However, common data augmentation methods are limited to perform complex data transformations, which would greatly help the classifiers to better generalize. A more powerful data augmentation method would be to use \textit{generative adversarial networks (GANs)}~\cite{goodfellow2014generative} to generate fake source domains with the style of target domain. Here, the main drawback is that the generated samples are representative only for the target domain. However, in multi-source case, we want the generated samples to be representative for all the domains we have at hand. In addition, style transfer needs to be performed between the target and each source domain; therefore, it is inconvenient.

In the field of remote sensing, each satellite image can be regarded as a domain. In our multi-source domain adaptation problem definition, we assume that each source and target domains have significantly different data distributions (see the real data in the first row of Fig.~\ref{fig:teaser}). Our method aims at finding a common representation for all the domains by \textit{standardizing} the samples belonging to each domain using GANs. As shown in Fig.~\ref{fig:teaser}, in a way, the standardized data could be considered as spectral interpolation across the domains. Adopting such a standardization strategy has two advantages. Firstly, in the training stage, it prevents the classifier from capturing the idiosyncrasies of each source domain. The classifier rather learns from the common representation. Secondly, since in the common representation the samples belonging to source domains and target domain have distributions close to each other, we expect the classifier trained on the standardized source domains to segment well the standardized target domain.

Standardizing multiple domains using GANs raises several challenges. Firstly, when training GANs, one needs real data so that the generator can generate fake data with the distribution that is as close as possible to the distribution of the real data. However, in our case, the standardized data do not exist. In other words, we wish to generate data without showing samples drawn from a similar distribution. Secondly, all the standardized domains need to have similar data distributions. Otherwise, the advantages mentioned above would be lost. Thirdly, the standardized data and the real data themselves must be semantically consistent. For example, when generating the standardized data, the method should not replace some objects by the others, add artificial objects, or remove some objects existing in the real data. Otherwise, the standardized data and the ground-truth for the real data would not match, and we could not train a model. Finally, the method should be efficient. If the number of networks and their structures are not kept as small as possible, depending on the number of domains, we could face with issues in terms of memory occupation and computational time.

In this work, we present novel StandardGAN, which overcomes all the aforementioned challenges. The main contributions are three fold. Firstly, we introduce the use of GANs in the context of data standardization. Secondly, we present a GAN that is able to generate data samples without providing it with data coming from the same or similar distribution. Finally, we propose to apply this multi-source domain adaptation solution to the semantic segmentation of Pl{\'e}iades data collected over several geographic locations.

\section{Related Work}\label{sec:related_work}
\paragraph{Adapting the classifier.} These methods aim at adapting the classifier to target domain. A common approach is to perform multi-task learning, where one of the tasks is to train a classifier from the source domain via common supervised learning approaches, and the other one is to align the features extracted from both source and target domains by adversarial training~\cite{hoffman2016fcns, tsai2018learning, huang2018domain}. A similar approach~\cite{deng2019large} has also been applied to remote sensing data (SpaceNet challenge~\cite{van2018spacenet}). Other approaches include self learning~\cite{zhang2018fully, zou2018domain}, using task-specific decision boundaries~\cite{saito2018maximum}, introducing new normalization~\cite{romijnders2019domain, pan2018two} or regularization methods~\cite{saito2017adversarial}, and adding specific loss functions for domain adaptation~\cite{zhu2018penalizing}.

\vspace{-2mm}
\paragraph{Adapting the inputs.} These methods, in general, try to perform image-to-image translation (I2I) or style transfer between domains to generate target stylized fake source data. The fake data are then used to train or to fine-tune the classifier. For example, CyCADA~\cite{hoffman2017cycada} uses CycleGAN~\cite{zhu2017unpaired} to generate target stylized fake source data. CycleGAN has also been applied to aerial images~\cite{benjdira2019unsupervised}. For the style transfer between satellite images, Tasar~\etal have recently introduced ColorMapGAN~\cite{tasar2019colormapgan} that learns to map each color of the source image to another one, and SemI2I~\cite{tasar2020semi2i} that switches the styles of the source and the target domains. To accomplish the same task, one can also consider using other I2I approaches in the computer vision community such as UNIT~\cite{liu2017unsupervised}, MUNIT~\cite{huang2018multimodal}, DRIT~\cite{lee2018diverse}, or common approaches like histogram matching~\cite{Gonzalez}.

\begin{figure*}
\centering
    \includegraphics[width=1\linewidth]{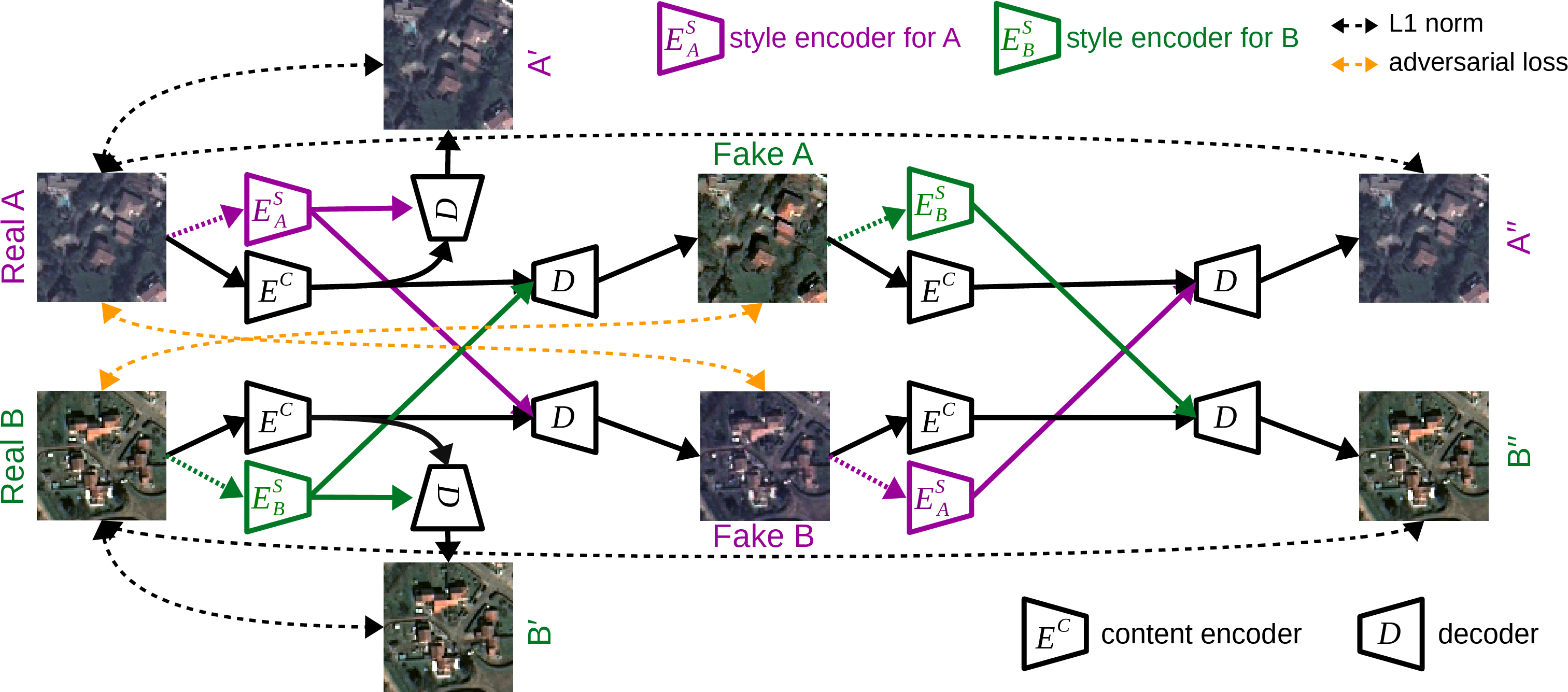}
\caption{Style transfer between two cities. In this example, there exists 2 style encoders, 1 content encoder, 1 decoder, and 1 discriminator.}
\label{fig:i2i}
\end{figure*}

\begin{figure}
\centering
    \includegraphics[width=1\linewidth]{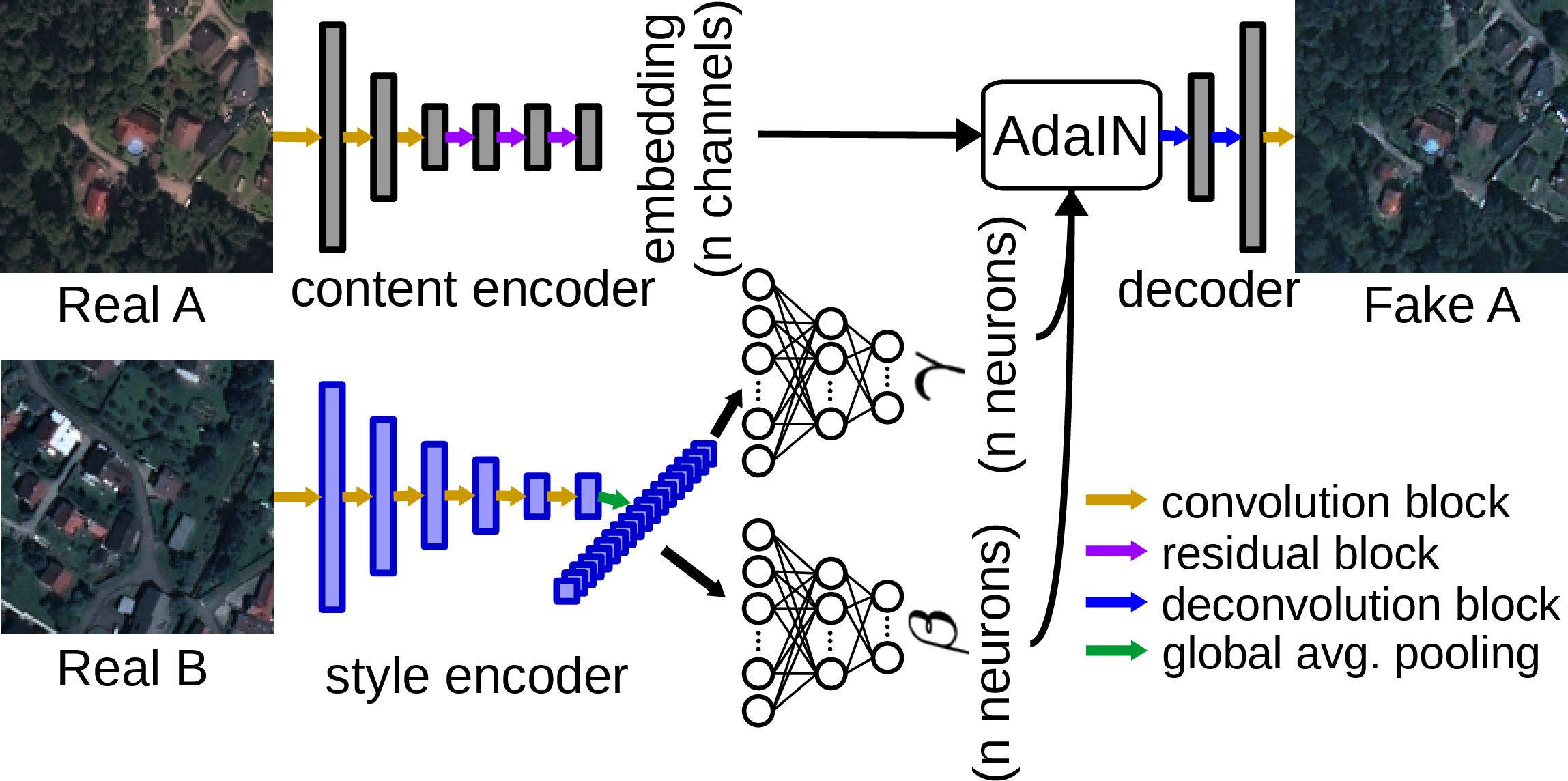}
\caption{Combining the content of one city with the style of another city.}
\label{fig:adaIN}
\end{figure}

\vspace{-2mm}
\paragraph{Multi-source domain adaptation (MDA).} The most straightforward approach would be to perform I2I between each source and target domains to stylize all of the source domains as target domain. However, this method is extremely cumbersome, because the training must be performed for each source domain and the target domain pair. In addition, the data distribution of each source domain is made similar to the distribution of only one domain (i.e., target domain). Instead, finding a common representation that is representative for all the domains is desired. Recently, specifically for MDA, a few methods focusing on image classification have been proposed~\cite{zhao2018adversarial, xu2018deep, peng2019moment}. However, it may not be possible to extend these works to semantic segmentation, as precisely structured output is required. To address the issue of MDA for semantic segmentation, Zhao~\etal have proposed MADAN~\cite{zhao2019multi}, which is an extension of CyCADA, but it is extremely compute-intensive. JCPOT~\cite{redko2018optimal} investigates optimal transport for MDA problem. Elshamli~\etal have recently proposed a method consisting in patch based networks~\cite{elshamli2019multisource}. However, since the network architectures are not fully convolutional, the method may not be suitable for classes requiring high precision such as buildings and roads.

\vspace{-2mm}
\paragraph{Data standardization. } In machine learning, one of the most commonly used data standardization approach is referred to as Z-score normalization and computed as:
\begin{equation}\label{eq:z_norm}
x^{\prime} = \frac{x - \mu}{\sigma},
\end{equation}
where $x$, $\mu$, $\sigma$ correspond to original data, mean value, and standard deviation. In addition, histogram equalization~\cite{Gonzalez} is also a common pre-processing step. However, these approaches do not take into account the contextual information, they just follow certain heuristics. One may also think of applying color constancy algorithms~\cite{agarwal2006overview} such as gray-world~\cite{buchsbaum1980spatial} and gamut~\cite{forsyth1990novel} approaches. These algorithms assume that colors of the objects are highly affected by the color of the illuminant and try to remove this effect.

\section{Method}
In this section, we first explain how to perform style transfer between two domains. We then describe how StandardGAN standardizes two domains. Finally, we detail how we extend StandardGAN to multi-domain case. 

StandardGAN consists of one content encoder, one decoder, one discriminator, and $n$ style encoders, where $n$ is the number of domains. Fig.~\ref{fig:i2i} illustrates the generator to perform style transfer between two domains. The discriminator performs multi-task learning as in StarGAN~\cite{choi2018stargan} by adding an auxiliary classifier on top of the discriminator of CycleGAN~\cite{zhu2017unpaired}. The first task allows the fake source and the target domains to have  as similar data distributions as possible, whereas the other task helps the discriminator to understand between which fake and real data it is discriminating. We provide detailed explanations for both tasks in \textit{style transfer} and \textit{classification loss} parts of the following sub-section.

\subsection{Style Transfer Between Two Domains}
We denote both domains by A and B. In the following, we explain the main steps that are required for style transfer between two domains.

\vspace{-2mm}
\paragraph{Style Transfer.} The goal of style transfer is to generate fake A with the style of B and fake B having a similar data distribution as real A. To perform style transfer, we use two types of encoders. One is domain agnostic content encoder, and the other one is domain specific style encoder. The content encoder is used to map the data into a common space, irrespective of which domain the data come from. On the other hand, the style encoder helps the decoder to generate output with the style of its specific domain. We use adaptive instance normalization (AdaIN)~\cite{huang2017arbitrary} to combine the content of A with the style of B (or vice versa). AdaIN is defined as:
\begin{equation}\label{eq:adaIN}
\text{AdaIN}( x, \gamma, \beta ) = \gamma \left( \frac{x - \mu(x)}{\sigma(x)} \right) + \beta,
\end{equation}
where $x$ is the activation of the content encoder's final convolutional layer, and $\gamma$ and $\beta$ correspond to the parameters that are learned by the style encoder. As can be seen in Eq.~\ref{eq:adaIN}, $\gamma$ and $\beta$ are used to scale and shift the activation, which results in changing the style of the output. After the activation is normalized by AdaIN, as depicted by Fig.~\ref{fig:adaIN}, it is fed to the decoder to generate the fake data.

In order to force real A and fake B, and real B and fake A to have as similar data distributions as possible, we compute and minimize an adversarial loss between them. We use the adversarial loss functions described in LSGAN~\cite{mao2017least}. The discriminator adversarial loss between real A and fake B (or real B and fake A) is defined as:
\begin{equation}
\begin{split}
\mathcal{L}_{adv{\_}D} = \mathbb{E}_{x \sim p(x)}[(D_{adv}(x) - 1)^2] ~+~ \\ \mathbb{E}_{y \sim p(y)}[(D_{adv}(G(y)))^2]
\end{split}
\end{equation}
where $\mathbb{E}$ denotes the expected value, $G$ and $D_{adv}$ stand for the generator and the adversarial output of the discriminator (the first task), and $x$ and $y$ correspond to data for both domains drawn from the distributions of $p(x)$ and $p(y)$. The generator adversarial loss is computed as:
\begin{equation}
\mathcal{L}_{adv{\_}G} = \mathbb{E}_{y \sim p(y)}[(D_{adv}(y) - 1)^2].
\end{equation}
The overall generator adversarial loss $\mathcal{L}_{adv{\_}G}$ and the discriminator adversarial loss $\mathcal{L}_{adv{\_}D}$ are calculated by simply summing the adversarial losses between real A and fake B, and real B and fake A. 

\vspace{-2mm}
\paragraph{Classification loss.}
To force real A and fake B, and real B and fake A to have similar styles, normally, we need two discriminators. One is used for discriminating between real A and fake B, and the other is responsible for distinguishing between real B and fake A. However, as mentioned in Sec.~\ref{sec:intro}, we want to keep the number of networks as small as possible to easily extend StandardGAN to multi-domain case. In order to use only one discriminator, we adopt the strategy explained in StarGAN~\cite{choi2018stargan}. Let us assume that A is the source and B is the target domain. We suppose that the labels of A and B are indicated by $c\_s$ and $c\_t$ (e.g., $c\_s = 0$ and $c\_t = 1$), and the image patch sampled from A is denoted by $x$. On top of the discriminator, we add a classifier. Both the discriminator and the generator have a role on this classifier. On the one hand, the discriminator wants the classifier to predict the label of A correctly. On the other hand, the generator tries to generate fake A in a way that the classifier predicts it as B. The classification loss for the discriminator is defined as:
\begin{equation}\label{eq:disc_cls_loss}
\mathcal{L}_{cls{\_}D} = \mathbb{E} [ -\text{log} D_{cls} (c\_s \mid x )],
\end{equation}
where $D_{cls} (c\_s \mid x )$ denotes the probability distribution over domain labels generated by $D$. By minimizing this function, $D$ learns from which domain $x$ come. The classification loss for the generator is computed as:
\begin{equation}\label{eq:gen_cls_loss}
\mathcal{L}_{cls{\_}G} = \mathbb{E} [ -\text{log} D_{cls} (c\_t \mid G(x)) ].
\end{equation}
Minimizing this function causes $D$ to label fake A ($G(x)$) as B. We sum the classification losses between real A and fake B, and real B and fake A to compute the overall domain classification losses $\mathcal{L}_{cls\_D}$ and $\mathcal{L}_{cls\_G}$. In the training stage, minimizing Eqs.~\ref{eq:disc_cls_loss} and \ref{eq:gen_cls_loss} allows the discriminator to understand whether it needs to distinguish between real A and fake B or between real B and fake A. As a result, the style transfer can be performed with only one discriminator. The classification loss is particularly useful when we extend StandardGAN to multi-domain adaptation case.

\vspace{-2mm}
\paragraph{Semantic Consistency.} As mentioned in Sec.~\ref{sec:intro}, it is crucial to perform the style transfer without spoiling the semantics of the real data. Otherwise, the fake data and the ground-truth for the real data would not overlap. Thus, they cannot be used to train a model. For this reason, our decoder is architecturally quite simple. It consists of only one convolution and two deconvolution blocks (see Fig.~\ref{fig:adaIN}). After scaling and shifting the content embedding of one domain with the AdaIN parameters learned by the style encoder from another domain, we directly decode the embedding, instead of adding further residual blocks. Moreover, we have additional constraints enforcing semantic consistency. As shown in Fig.~\ref{fig:i2i}, after we generate fake A with the style of B and fake B with the style of real A, we switch the styles once again to obtain A$''$ and B$''$. In an ideal case, A and A$''$, and B and B$''$ must be the same. Hence, we minimize the cross reconstruction loss $\mathcal{L}_{cross}$ that is the sum of L1 norms between A and A$''$, and between B and B$''$. Similarly, when we combine the content information of a domain with its own style information, we should be reconstructing itself (see A$'$ and B$'$ in Fig.~\ref{fig:i2i}). We also minimize the self reconstruction loss $\mathcal{L}_{self}$, which is computed by summing the L1 norms between A and A$'$, and between B and B$'$.

\vspace{-2mm}
\paragraph{Training.} The overall generator loss is calculated as:
\begin{equation}\label{eq:g_loss}
\mathcal{L}_{G} = \lambda_{1} \mathcal{L}_{cross} + \lambda_{2} \mathcal{L}_{self} + \lambda_{3} \mathcal{L}_{cls\_{G}} + \lambda_{4} \mathcal{L}_{adv\_G}, 
\end{equation}
where $\lambda_{1}, \lambda_{2}, \lambda_{3}$, and $\lambda_{4}$ denote the weights for the individual losses. The discriminator loss is defined as:
\begin{equation}\label{eq:d_loss}
\mathcal{L}_{D} = \lambda_{3} \mathcal{L}_{cls\_{D}} + \lambda_{4} \mathcal{L}_{adv\_D}.
\end{equation}
We minimize $\mathcal{L}_{G}$ and $\mathcal{L}_{D}$ simultaneously. 

As can be seen in Fig.~\ref{fig:adaIN}, to generate fake data, content encoder, decoder, and the AdaIN parameters learned by the style encoder of the other domain are required. The issue is that the style encoder produces different AdaIN parameters for each image patch depending on the context of the patch. For instance, we cannot expect patches from a forest and an industrial area to have similar parameters, because they have different styles. For each domain, to capture the global AdaIN parameters, we first initialize domain specific $\gamma$ and $\beta$ parameters with zeros. We then propose to update them in each training iteration as:
\begin{equation}\label{eq:moving_avg}
p = 0.95 \times p + 0.05 \times p\_\text{current},
\end{equation}
where $p$ is the global domain specific AdaIN parameter (i.e., $\gamma$ or $\beta$) and $p\_\text{current}$ is the parameter from the current training patch. After a sufficiently long training process, Eq.~\ref{eq:moving_avg} estimates the global AdaIN parameters for each domain. These estimations can then be used in the test stage.

\begin{figure}
\centering
    \includegraphics[width=1\linewidth]{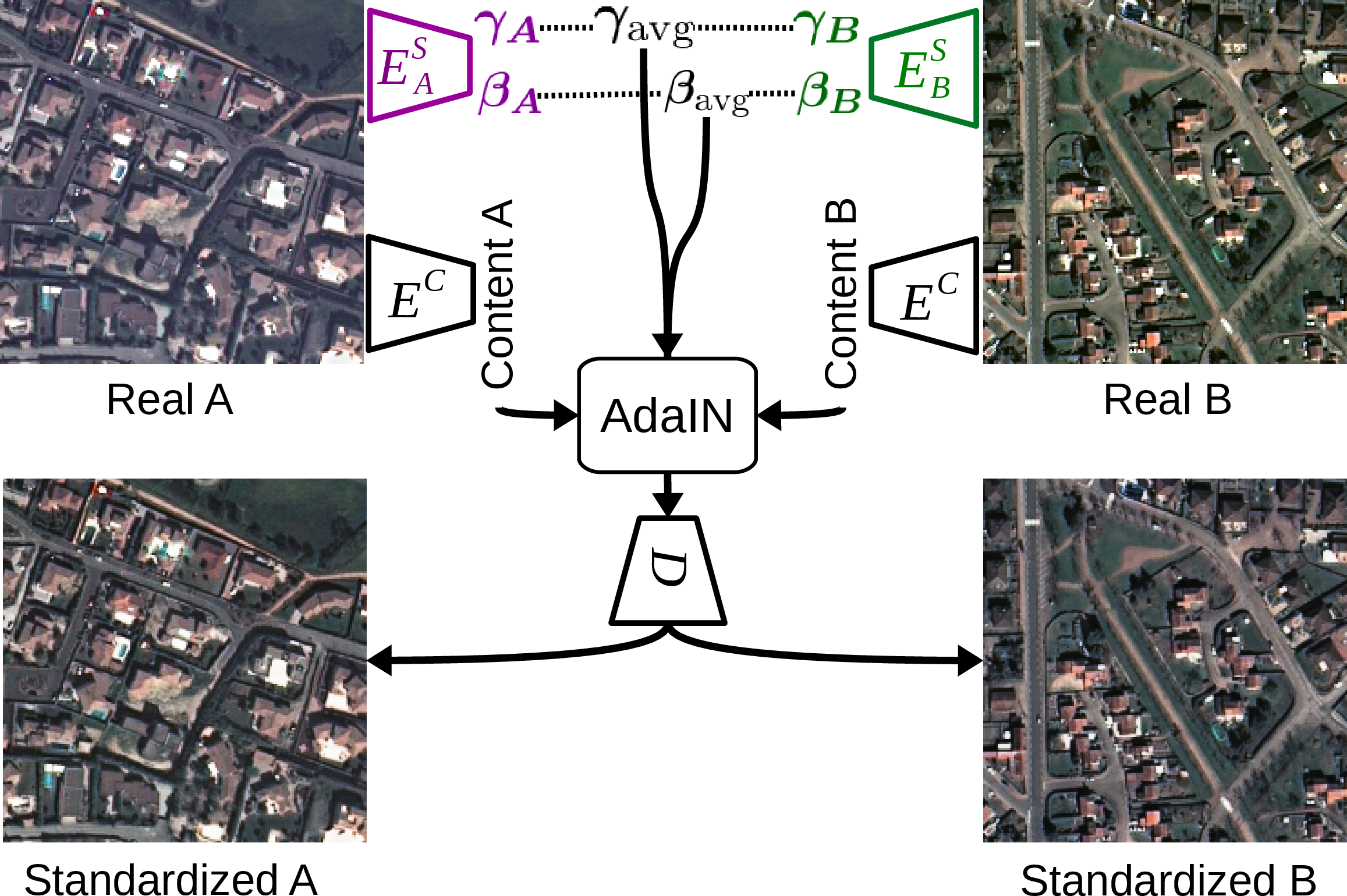}
\caption{Standardizing two domains. Dashed lines correspond to arithmetic average.}
\label{fig:standardization_two_domains}
\end{figure}

\begin{figure}
\centering
    \includegraphics[width=1\linewidth]{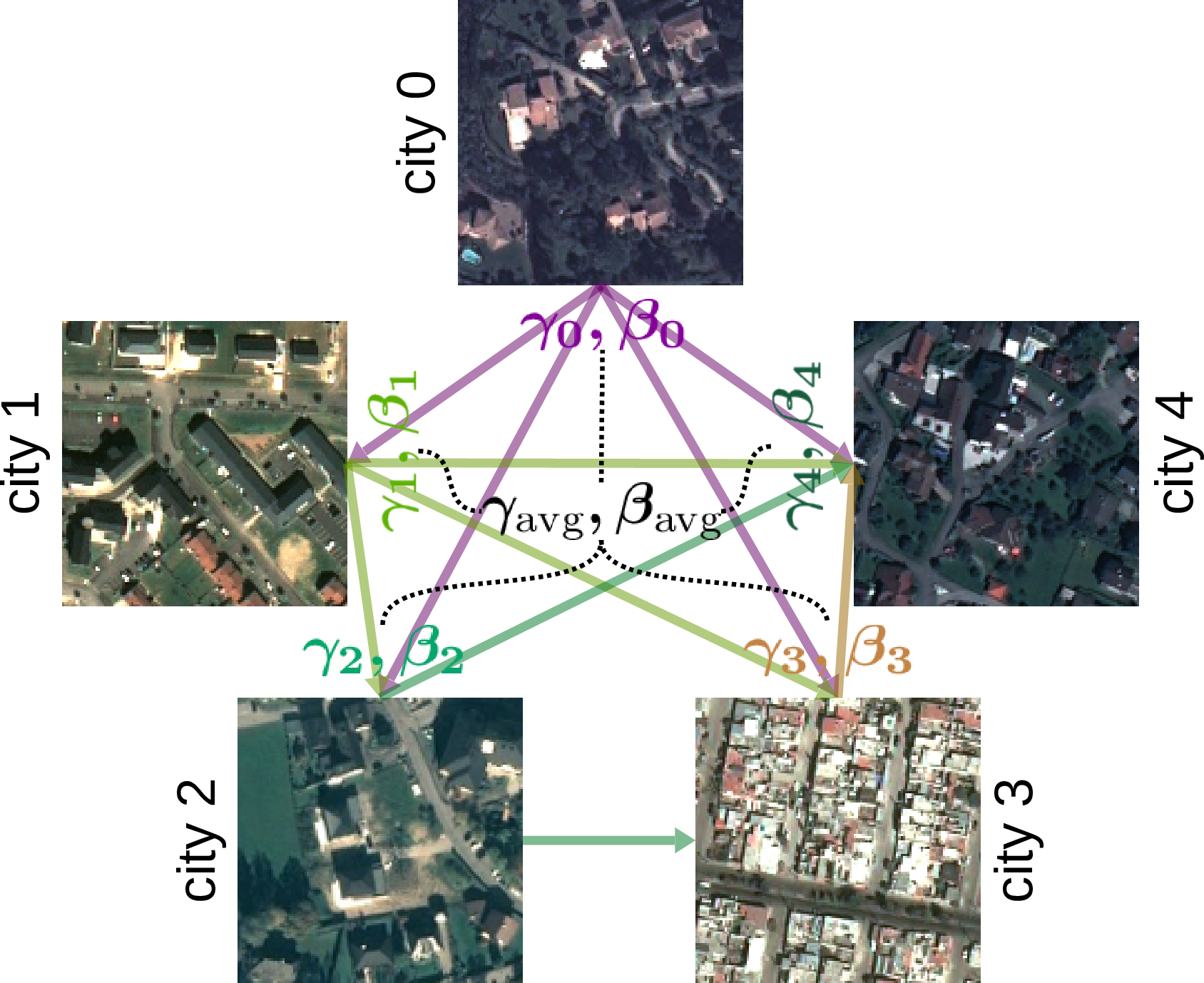}
\caption{Standardizing multiple domains. Solid arrows represent adaptation between two domains. Dashed lines correspond to arithmetic average. $\gamma_{\text{avg}}$ and $\beta_{\text{avg}}$ are used for standardization.}
\label{fig:standardization_multi_domains}
\end{figure}

\begin{algorithm}
\SetAlgoLined
\SetKwInOut{Input}{input}\SetKwInOut{Output}{output}
create 1 content encoder, 1 decoder, and 1 discrim.

\ForEach{domain}{
	init. domain specific AdaIN params. with zeros

	create a domain specific style encoder
}

\ForEach{training iteration}{

	$\mathcal{L}_{G} \gets 0$, $\mathcal{L}_{D} \gets 0$ \tcp*[l]{G and D losses}	

	\For{$i \gets 0$ \KwTo ($\#$ of domains - 1)}{
		\For{$j\gets (i + 1)$ \KwTo ($\#$ of domains - 1)}{
					
			$\mathcal{L}_{G\_} \gets $ G loss between dom. $i \& j$ (Eq.~\ref{eq:g_loss})
			
			$\mathcal{L}_{D\_} \gets $ D loss between dom. $i \& j$ (Eq.~\ref{eq:d_loss})
			
			$\mathcal{L}_{G} \gets \mathcal{L}_{G} + \mathcal{L}_{G\_}$
			
			$\mathcal{L}_{D} \gets \mathcal{L}_{D} + \mathcal{L}_{D\_}$
		}
    }
    
    backprop. $\mathcal{L}_{G}$ and $\mathcal{L}_{D}$, $\mathcal{L}_{G} \gets 0$, $\mathcal{L}_{D} \gets 0$
    
    \ForEach{domain}{
		update dom. spec. AdaIN params. via Eq.~\ref{eq:moving_avg}
    }

	avg. AdaIN params. $\gets$ arithmetic average of domain specific AdaIN parameters  
    
}
\caption{The pseudocode for StandardGAN.}\label{alg:standardization_multi}
\end{algorithm}

\subsection{StandardGAN for Image Standardization}
As mentioned previously, the domain agnostic content encoder learns to map domains into a common space. To generate target stylized fake source data, the content embedding extracted by the content encoder from the source domain is normalized with the global AdaIN parameters of the target domain. The normalized embedding is then given to the decoder to generate the fake data. We have discovered that instead of normalizing the embedding with the AdaIN parameters for one of the domains, if we normalize it with the arithmetic average of the global AdaIN parameters of both domains, StandardGAN learns to generate standardized data. The standardization process for two domains is depicted in Fig.~\ref{fig:standardization_two_domains}. As shown in the figure, real A and real B have considerably different data distributions. On the other hand, standardized A and standardized B look quite similar, and their data distributions are somewhere between the data distributions of real A and real B.

To standardize multiple domains, we propose Alg.~\ref{alg:standardization_multi}. In multi-domain case, $c\_s$ and $c\_t$ in Eqs.~\ref{eq:disc_cls_loss} and \ref{eq:gen_cls_loss} can range between 0 and $n$ - 1, where $n$ is the number of domains. As shown in Fig.~\ref{fig:standardization_multi_domains}, we perform adaptation between each pair of domains. We then take the average of the global AdaIN parameters of each domain and use the average to normalize the embeddings extracted by the content encoder from all the domains. We finally decode the normalized embeddings via the decoder to generate the standardized data.

\section{Experiments}

\begin{table}
\centering
\caption{The data set.}
\label{table:data_stats}
\begin{tabular}{c|c|c|c|c}
\hline
\multirow{2}{*}{\textbf{City (Country)}} & \multicolumn{3}{c|}{\textbf{Class percentages ($\%$)}} & \textbf{Area} \\ 
\cline{2-4}
                                         & \textbf{building} & \textbf{road} & \textbf{tree}      & \textbf{(km$^2$)}  \\
\hline
Bad Ischl (AT)                           &  5.51             &  6.0          & 35.38              &  27.71         \\
Salzburg Stadt (AT)                      &  9.44             &  8.69         & 23.88              & 134.71         \\
Villach (AT)                             &  9.26             & 10.63         & 19.91              &  43.59         \\
Lienz (AT)                               &  6.96             &  8.16         & 15.37              &  28.38         \\
Sankt P{\"o}lten (AT)                    &  6.68             &  6.39         & 25.13              &  87.17         \\
Bourges (FR)                             &  9.81             & 10.52         & 14.83              &  72.20         \\
Lille (FR)                               & 18.36             & 12.71         & 15.40              & 117.58         \\
Vaduz (LI)                               &  3.57             & 4.30          & 33.69              &  96.08         \\
\hline
\end{tabular}
\end{table}

In our experiments, we use Pl{\'e}iades images captured from 5 cities in Austria, 2 cities in France, and 1 city in Liechtenstein. The spectral channels consist of red, green, and blue bands. The spatial resolution has been reduced to 1 m by the data set providers. The annotations for building, road, and tree classes have been provided~\footnote{The authors would like to thank LuxCarta Technology for providing the annotated data that enabled us to conduct this research.}. Table~\ref{table:data_stats} reports, for each city, the name of the city, percentage of the pixels belonging to each class, and the total covered area. 

We have two experimental setups. In the first experiment, we use the images from Salzburg Stadt, Villach, Lienz, and Sankt P{\"o}lten for training and the image from Bad Ischl for test. In the second experiment, we choose Salzburg Stadt, Villach, Bourges, and Lille as the training cities and Vaduz as the test city. In the first experiment, we want to observe how well our method generalize to a new city from the same country. On the other hand, the goal of the second experiment is to investigate the generalization abilities of our approach when training and test data come from different countries. Let us also remark that, as confirmed by Table~\ref{table:data_stats}, classes in the test cities (i.e., Bad Ischl and Vaduz) are highly imbalanced, which makes the domain adaptation problem even more difficult. For example, in both cases, the number of pixels labeled as tree is significantly larger than the number of pixels labeled as building and road. 

\begin{figure}
\centering
\subfigure[Histograms for the real data]{%
    \includegraphics[width=\linewidth]{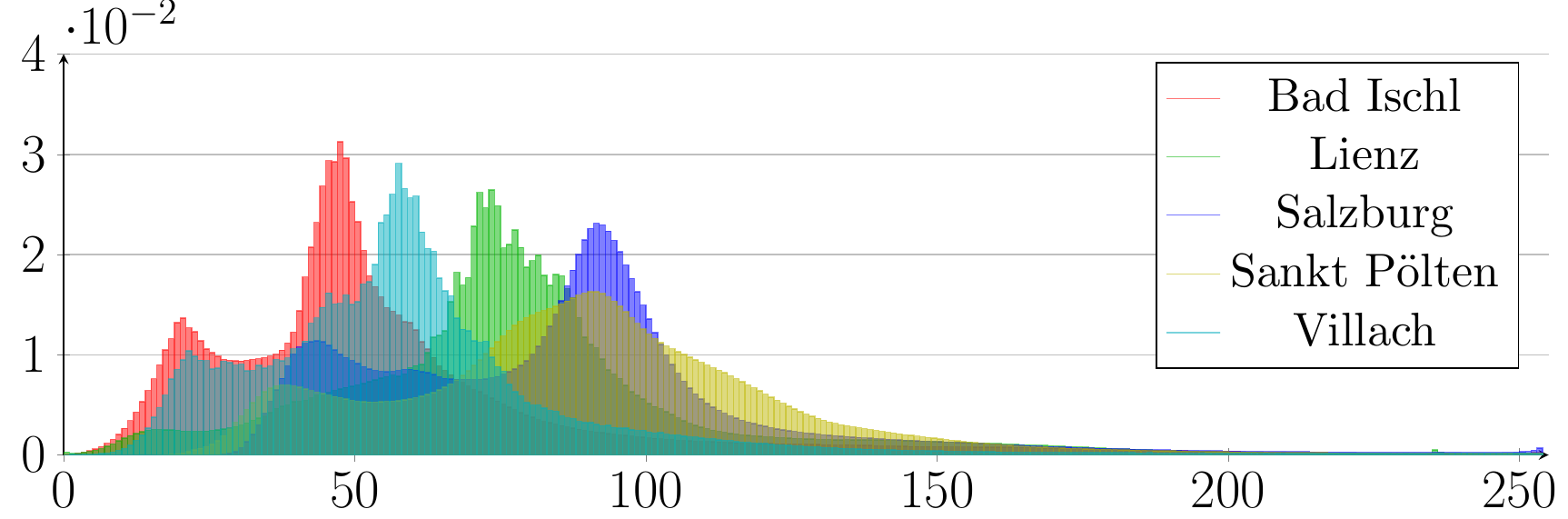}}
\subfigure[Histograms for the standardized data generated by StandardGAN]{%
    \includegraphics[width=\linewidth]{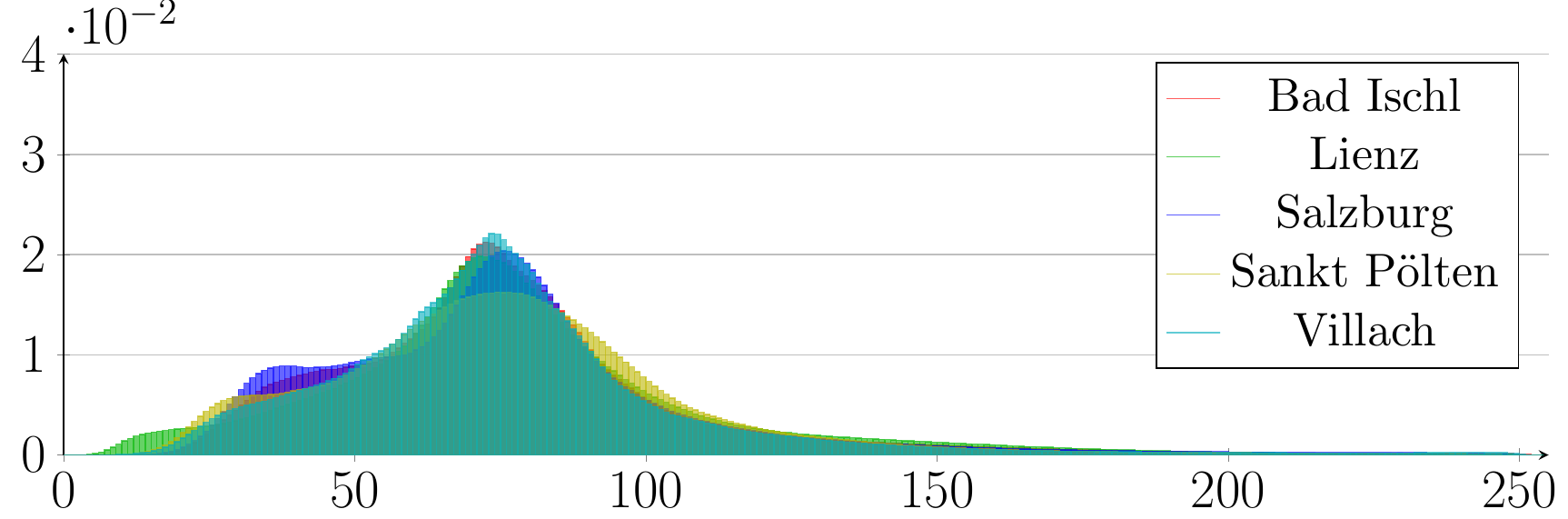}}
\caption{Histograms for green band of the cities used in the first experiment. (a) Before standardization, (b) After standardization.}
\label{fig:histograms}
\end{figure}

\begin{figure*}
\centering
\begin{tabular}{p{0.1em}c|c@{\hspace{0.15em}}c@{\hspace{0.15em}}c@{\hspace{0.15em}}c@{\hspace{0.15em}}c@{\hspace{0.15em}}}

\rotatebox[origin=c]{90}{Bad Ischl (0)}&
\raisebox{-.5\height}{\frame{\includegraphics[width=0.155\linewidth]{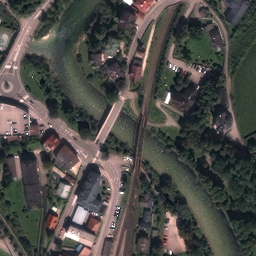}}} &
\raisebox{-.5\height}{\frame{\includegraphics[width=0.155\linewidth]{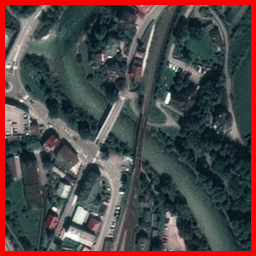}}} &
\raisebox{-.5\height}{\frame{\includegraphics[width=0.155\linewidth]{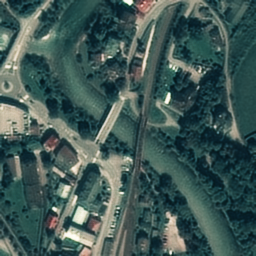}}} &
\raisebox{-.5\height}{\frame{\includegraphics[width=0.155\linewidth]{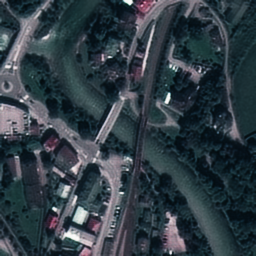}}} &
\raisebox{-.5\height}{\frame{\includegraphics[width=0.155\linewidth]{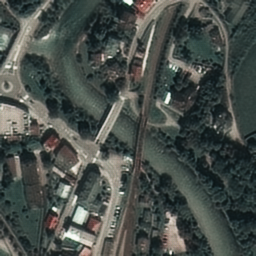}}} &
\raisebox{-.5\height}{\frame{\includegraphics[width=0.155\linewidth]{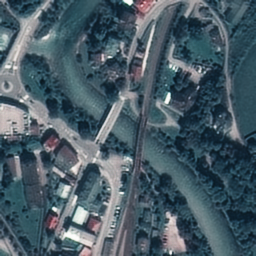}}}
\\[3.7em]
\rotatebox[origin=c]{90}{Salzburg Stadt (1)}&
\raisebox{-.5\height}{\frame{\includegraphics[width=0.155\linewidth]{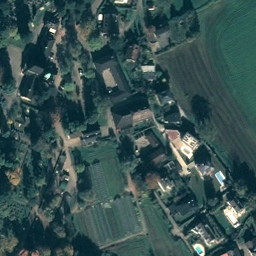}}} &
\raisebox{-.5\height}{\frame{\includegraphics[width=0.155\linewidth]{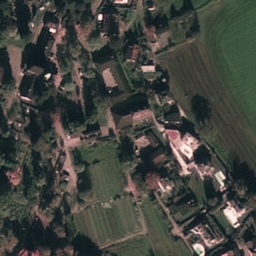}}} &
\raisebox{-.5\height}{\frame{\includegraphics[width=0.155\linewidth]{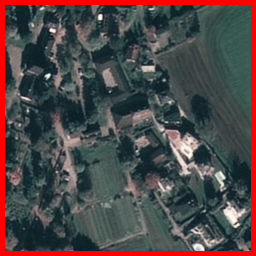}}} &
\raisebox{-.5\height}{\frame{\includegraphics[width=0.155\linewidth]{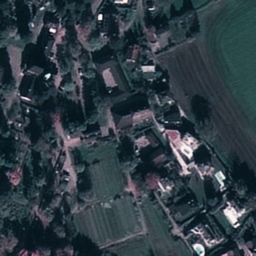}}} &
\raisebox{-.5\height}{\frame{\includegraphics[width=0.155\linewidth]{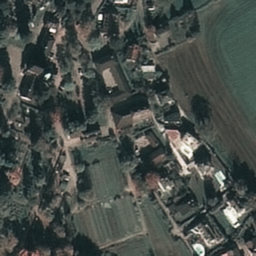}}} &
\raisebox{-.5\height}{\frame{\includegraphics[width=0.155\linewidth]{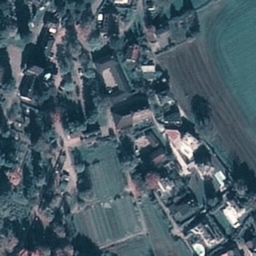}}}
\\[3.7em]
\rotatebox[origin=c]{90}{Villach (2)}&
\raisebox{-.5\height}{\frame{\includegraphics[width=0.155\linewidth]{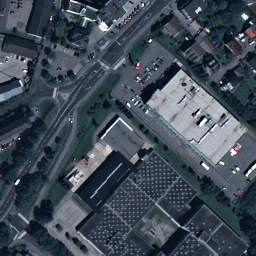}}} &
\raisebox{-.5\height}{\frame{\includegraphics[width=0.155\linewidth]{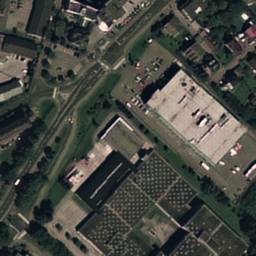}}} &
\raisebox{-.5\height}{\frame{\includegraphics[width=0.155\linewidth]{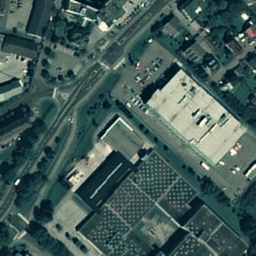}}} &
\raisebox{-.5\height}{\frame{\includegraphics[width=0.155\linewidth]{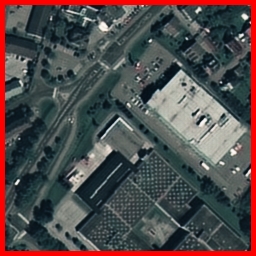}}} &
\raisebox{-.5\height}{\frame{\includegraphics[width=0.155\linewidth]{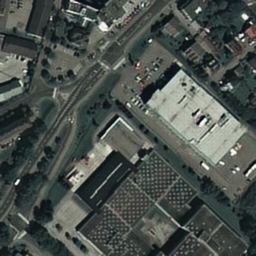}}} &
\raisebox{-.5\height}{\frame{\includegraphics[width=0.155\linewidth]{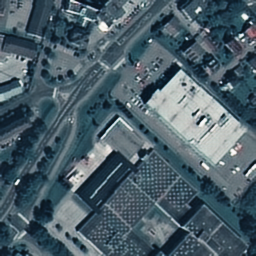}}}
\\[3.7em]
\rotatebox[origin=c]{90}{Lienz (3)}&
\raisebox{-.5\height}{\frame{\includegraphics[width=0.155\linewidth]{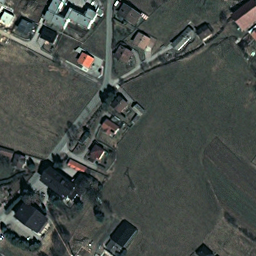}}} &
\raisebox{-.5\height}{\frame{\includegraphics[width=0.155\linewidth]{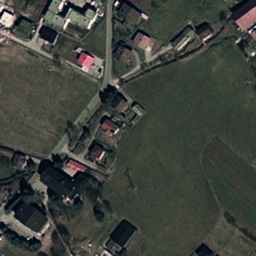}}} &
\raisebox{-.5\height}{\frame{\includegraphics[width=0.155\linewidth]{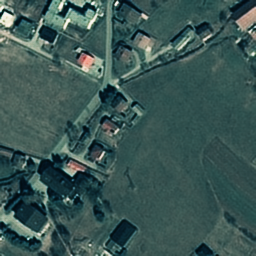}}} &
\raisebox{-.5\height}{\frame{\includegraphics[width=0.155\linewidth]{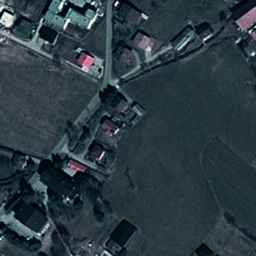}}} &
\raisebox{-.5\height}{\frame{\includegraphics[width=0.155\linewidth]{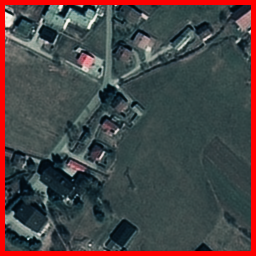}}} &
\raisebox{-.5\height}{\frame{\includegraphics[width=0.155\linewidth]{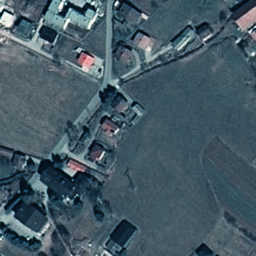}}}
\\[3.7em]
\rotatebox[origin=c]{90}{Sankt P{\"o}lten (4)}&
\raisebox{-.5\height}{\frame{\includegraphics[width=0.155\linewidth]{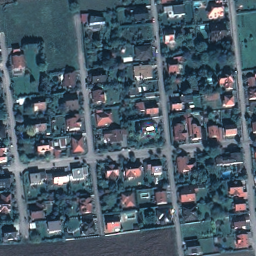}}} &
\raisebox{-.5\height}{\frame{\includegraphics[width=0.155\linewidth]{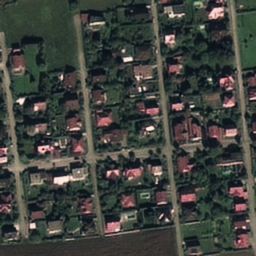}}} &
\raisebox{-.5\height}{\frame{\includegraphics[width=0.155\linewidth]{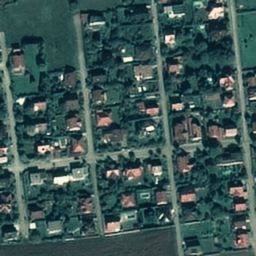}}} &
\raisebox{-.5\height}{\frame{\includegraphics[width=0.155\linewidth]{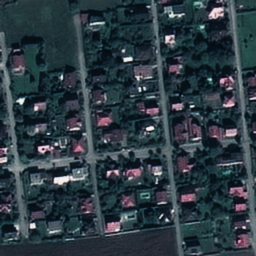}}} &
\raisebox{-.5\height}{\frame{\includegraphics[width=0.155\linewidth]{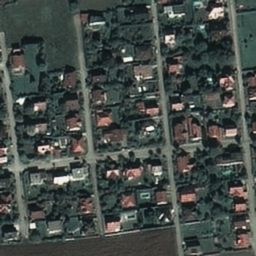}}} &
\raisebox{-.5\height}{\frame{\includegraphics[width=0.155\linewidth]{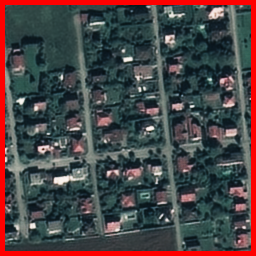}}}
\\[4em]
\end{tabular}
\caption{Real data used in the first experiment and the outputs generated by StandardGAN. Left column: the real data. Matrix on the right: The standardized data are highlighted by red bounding boxes. The rest of the cells depict the $i^{th}$ domain with the style of $j^{th}$ domain. The domain ids are indicated inside parentheses.}
\label{fig:fake_exp1}
\end{figure*}

\begin{table}
\centering
\caption{Training time of StandardGAN for both experiments.}
\label{table:training_time}
\begin{tabular}{cccc}
\hline
\textbf{GPU} & \textbf{Exp.} & \textbf{$\#$ of patches} & \textbf{Tr. time (secs.)} \\
\hline 
Nvidia Tesla & 1                & 5712                  &  6077.82                  \\
V100 SMX2    & 2                & 8226                  &  9929.52                  \\
\hline
\end{tabular}
\end{table}

\begin{table}
\centering
\caption{IoU scores for Bad Ischl (the first experiment).}
\label{table:ious_exp1}
\begin{tabular}{p{0.1em}p{0.1em}c@{\hspace{0.8em}}c@{\hspace{0.8em}}c@{\hspace{0.8em}}c@{\hspace{0.8em}}c@{\hspace{0.8em}}}
\hline
& & \textbf{Method} & \textbf{building} & \textbf{road} & \textbf{tree} & \textbf{Overall} \\
\hline

& & U-net           & 45.36             & 18.81         & \textbf{82.43}& 48.87            \\  
\multirow{4}{*}{\rotatebox[origin=c]{90}{U-net on}} &  
\multirow{4}{*}{\rotatebox[origin=c]{90}{data by}}   
& Gray-world        & 49.39             & 42.25         & 66.31         & 52.65            \\
& & Hist. Equaliz.  & 45.33             & 39.07         & 73.03         & 52.48            \\
& & Z-score norm.   & 51.22             & 46.56         & 77.62         & 58.47            \\
& & StandardGAN     & \textbf{56.41}    & \textbf{50.26}& 80.59         & \textbf{62.42}   \\
\hline
\end{tabular}
\end{table}

\begin{table}
\centering
\caption{IoU scores for Vaduz (the second experiment).}
\label{table:ious_exp2}
\begin{tabular}{p{0.1em}p{0.1em}c@{\hspace{0.8em}}c@{\hspace{0.8em}}c@{\hspace{0.8em}}c@{\hspace{0.8em}}c@{\hspace{0.8em}}}
\hline
& & \textbf{Method} & \textbf{building} & \textbf{road} & \textbf{tree} & \textbf{Overall} \\
\hline

& & U-net           & 29.83          & 26.42          & 46.25          & 34.16 \\  
\multirow{4}{*}{\rotatebox[origin=c]{90}{U-net on}}   &  
\multirow{4}{*}{\rotatebox[origin=c]{90}{data by}}   
& Gray-world        & 27.95          & 31.13          & 36.65          & 31.91 \\
& & Hist. Equaliz.  & 21.21          & 19.19          & 51.93          & 30.78 \\
& & Z-score norm.   & 29.94          & 29.87          & 41.98          & 33.93\\
& & StandardGAN     & \textbf{54.86} & \textbf{42.43} & \textbf{63.09} & \textbf{53.46} \\
\hline
\end{tabular}
\end{table}

\begin{figure*}
\centering
\begin{tabular}{p{0.1em}c@{\hspace{0.4em}}c@{\hspace{0.4em}}c@{\hspace{0.4em}}c@{\hspace{0.4em}}c@{\hspace{0.4em}}}

& Salzburg Stadt & Villach & Bourges & Lille & Vaduz \\
\rotatebox[origin=c]{90}{Real Data}&
\raisebox{-.5\height}{\frame{\includegraphics[width=0.18\linewidth]{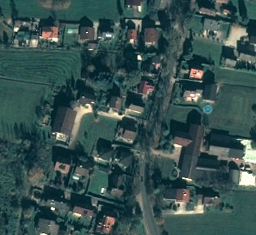}}} &
\raisebox{-.5\height}{\frame{\includegraphics[width=0.18\linewidth]{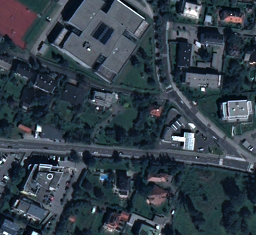}}} &
\raisebox{-.5\height}{\frame{\includegraphics[width=0.18\linewidth]{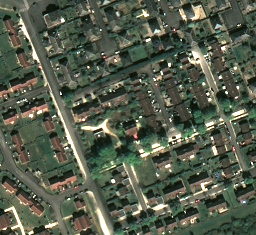}}} &
\raisebox{-.5\height}{\frame{\includegraphics[width=0.18\linewidth]{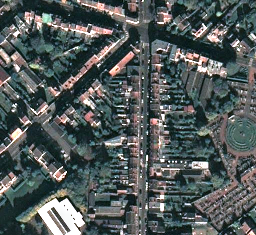}}} &
\raisebox{-.5\height}{\frame{\includegraphics[width=0.18\linewidth]{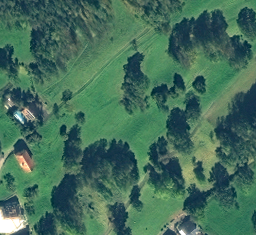}}}
\\[4.15em]
\rotatebox[origin=c]{90}{Standardized Data}&
\raisebox{-.5\height}{\frame{\includegraphics[width=0.18\linewidth]{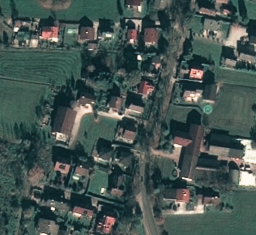}}} &
\raisebox{-.5\height}{\frame{\includegraphics[width=0.18\linewidth]{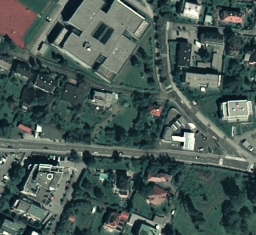}}} &
\raisebox{-.5\height}{\frame{\includegraphics[width=0.18\linewidth]{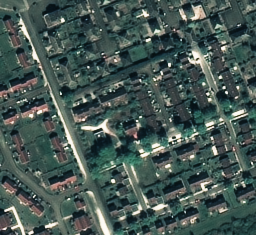}}} &
\raisebox{-.5\height}{\frame{\includegraphics[width=0.18\linewidth]{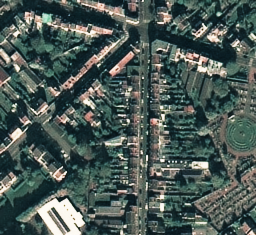}}} &
\raisebox{-.5\height}{\frame{\includegraphics[width=0.18\linewidth]{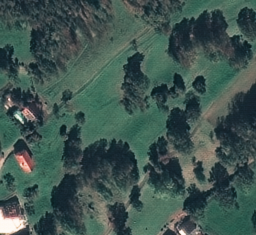}}}
\\[4.7em]
\end{tabular}
\caption{Real cities used in the second experiment, and the standardized data generated by StandardGAN.}
\label{fig:fake_exp6}
\end{figure*}

\begin{figure*}
\centering
\begin{tabular}{p{0.1em}c@{\hspace{0.4em}}c@{\hspace{0.4em}}c@{\hspace{0.4em}}c@{\hspace{0.4em}}}
& Real Data & Ground-Truth & U-net & Our framework\\
\rotatebox[origin=c]{90}{Bad Ischl} &
\raisebox{-.5\height}{\frame{\includegraphics[width=0.23\linewidth]{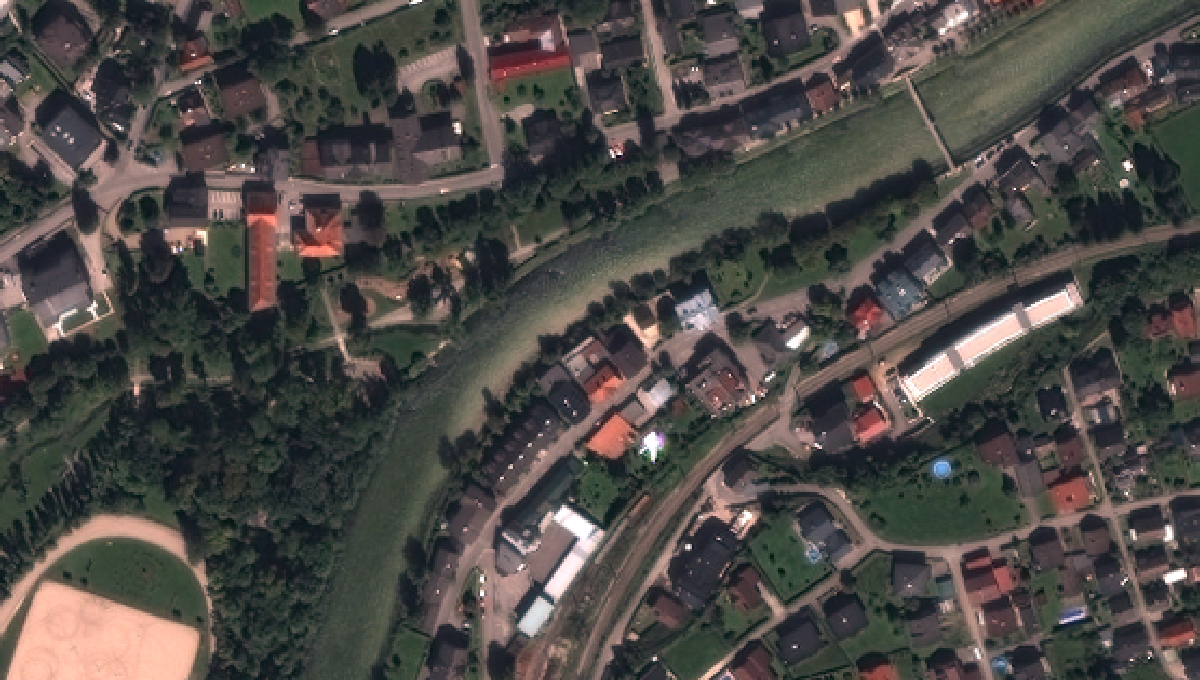}}} &
\raisebox{-.5\height}{\frame{\includegraphics[width=0.23\linewidth]{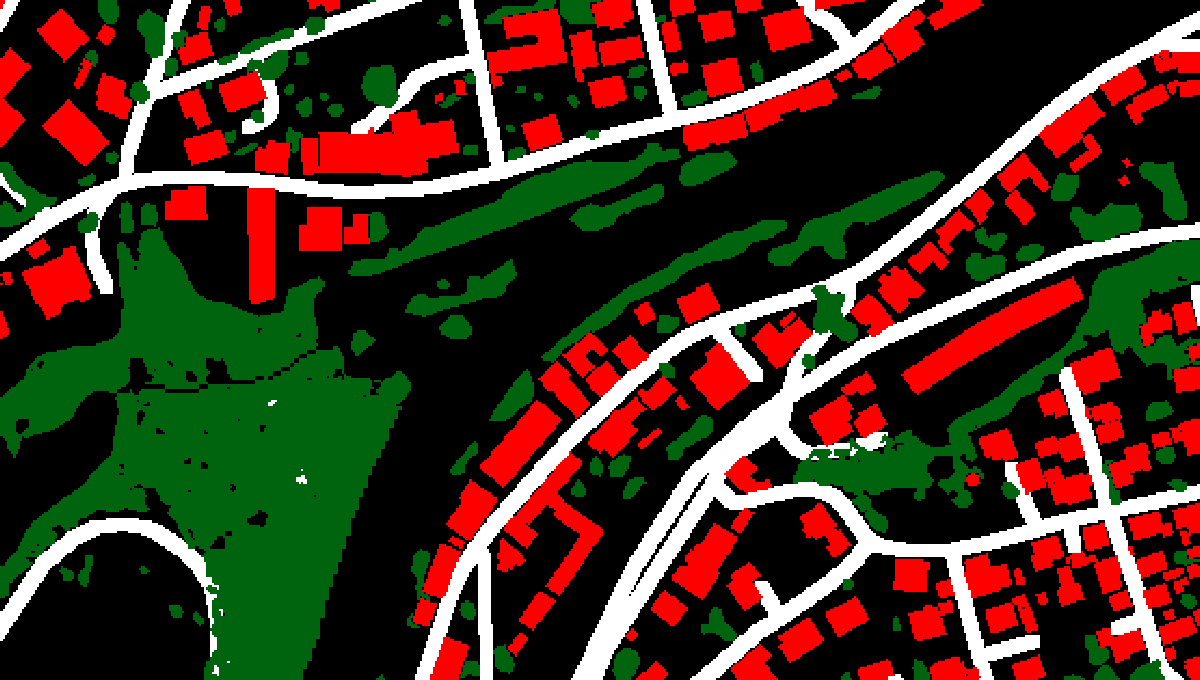}}} &
\raisebox{-.5\height}{\frame{\includegraphics[width=0.23\linewidth]{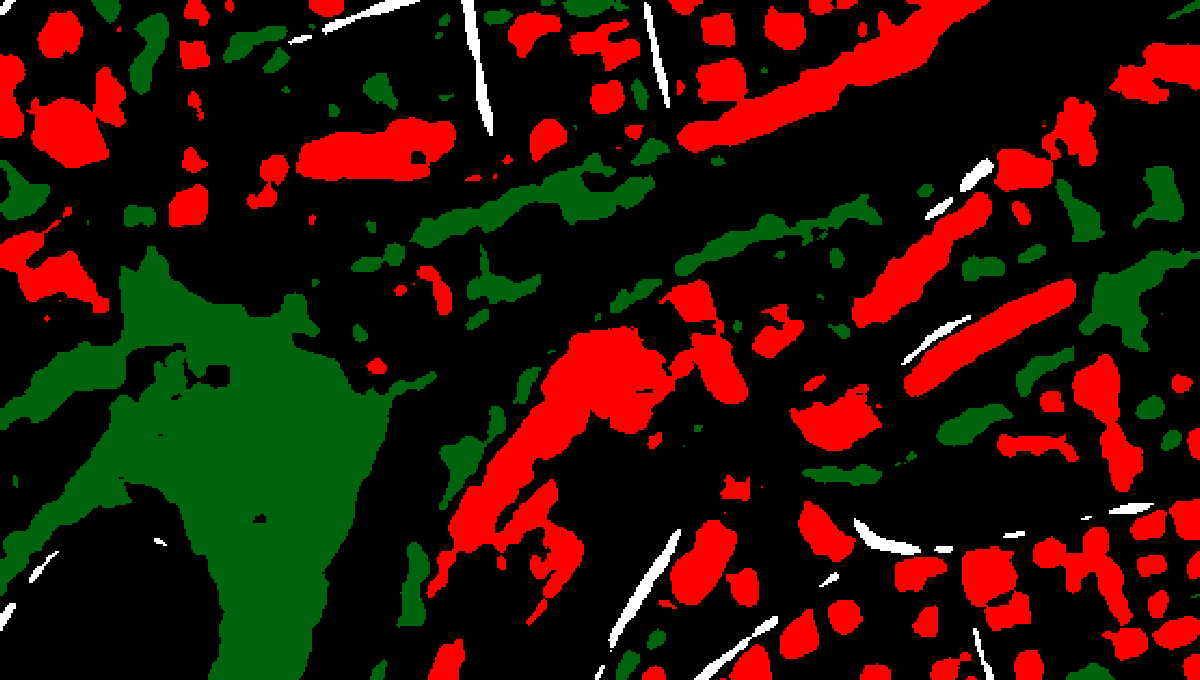}}} &
\raisebox{-.5\height}{\frame{\includegraphics[width=0.23\linewidth]{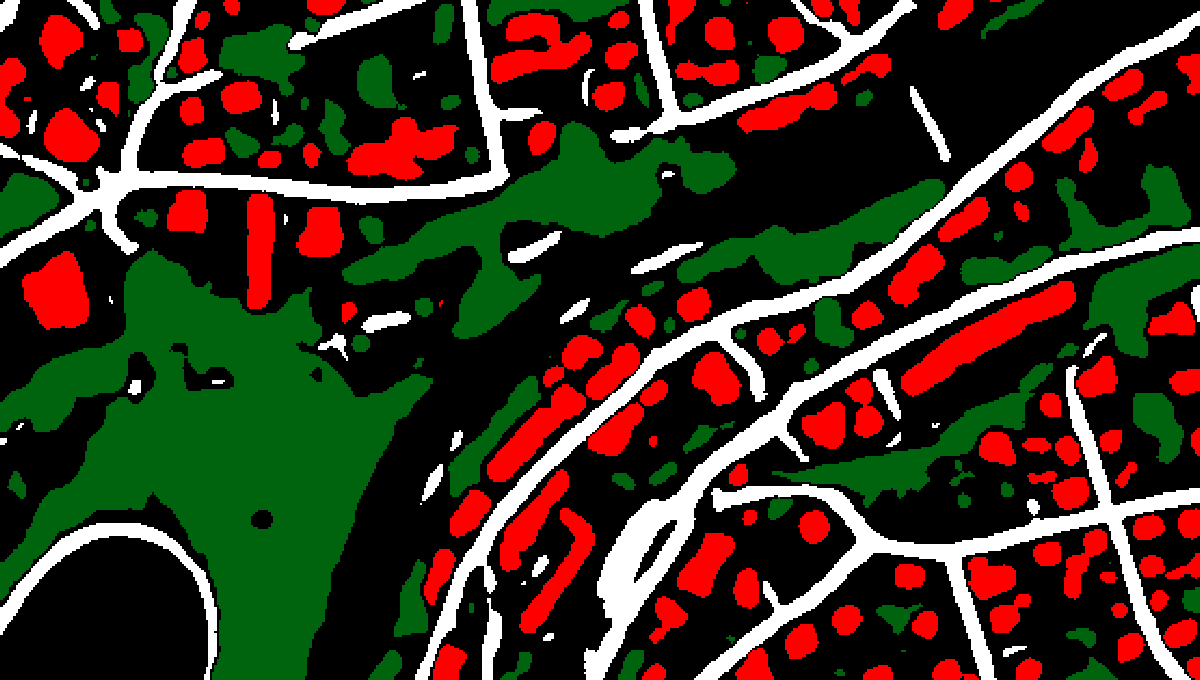}}}
\\[3.2em]
\rotatebox[origin=c]{90}{Vaduz} &
\raisebox{-.5\height}{\frame{\includegraphics[width=0.23\linewidth]{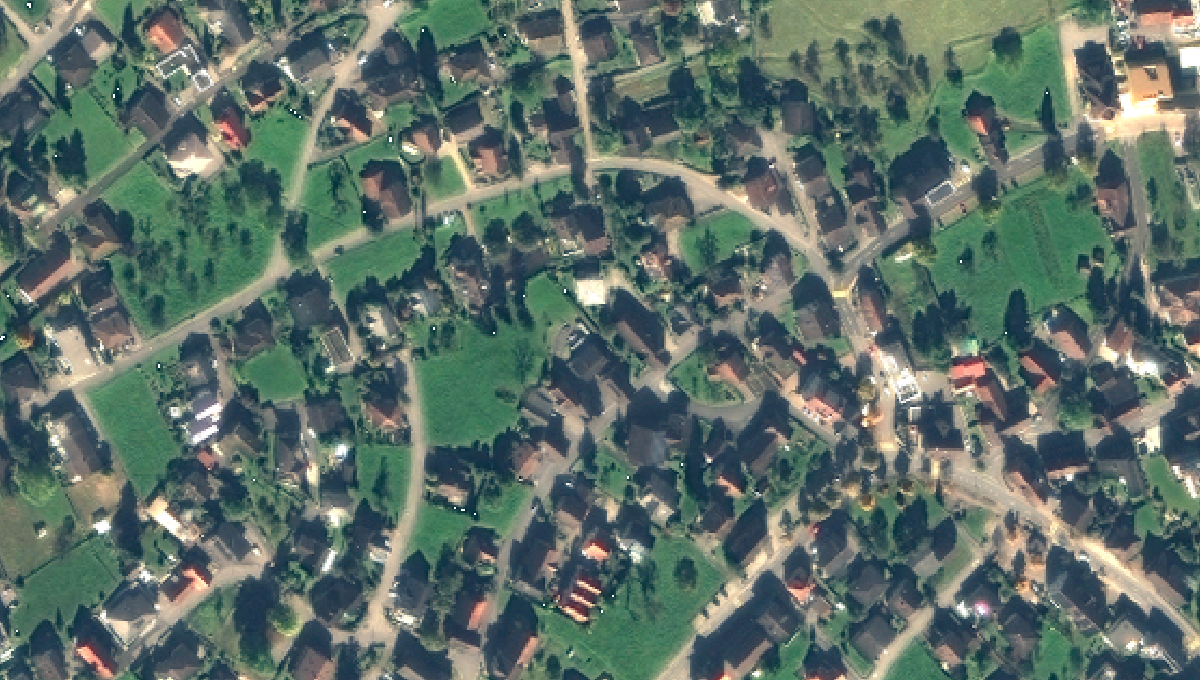}}} &
\raisebox{-.5\height}{\frame{\includegraphics[width=0.23\linewidth]{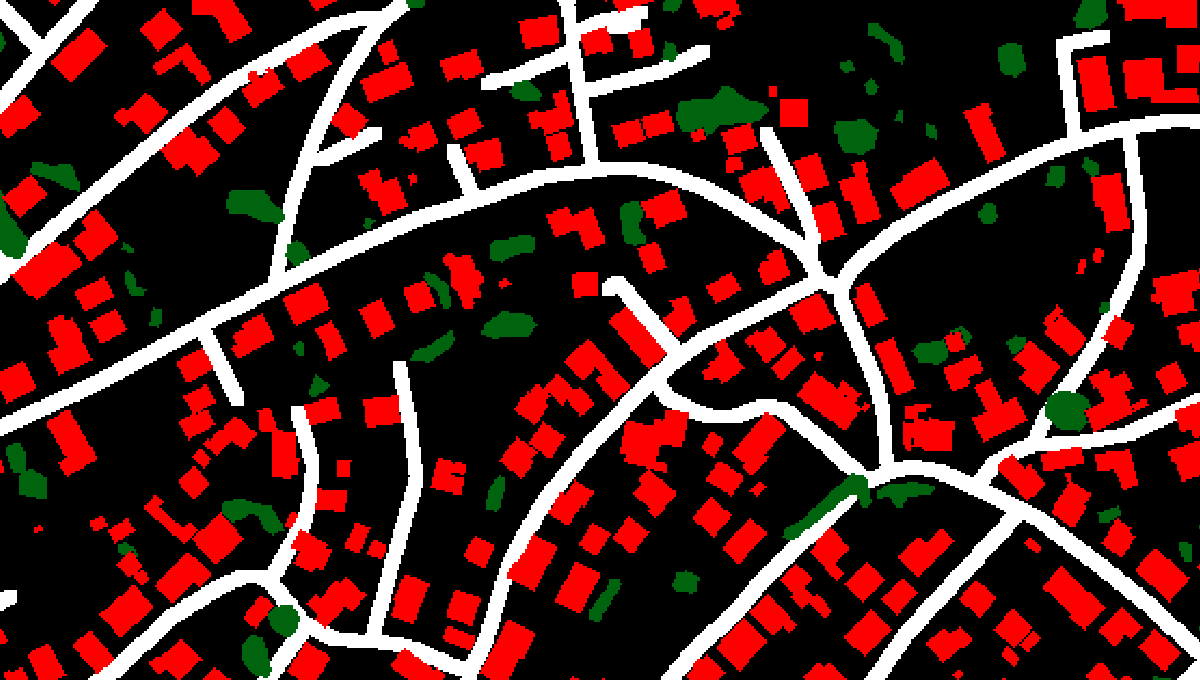}}} &
\raisebox{-.5\height}{\frame{\includegraphics[width=0.23\linewidth]{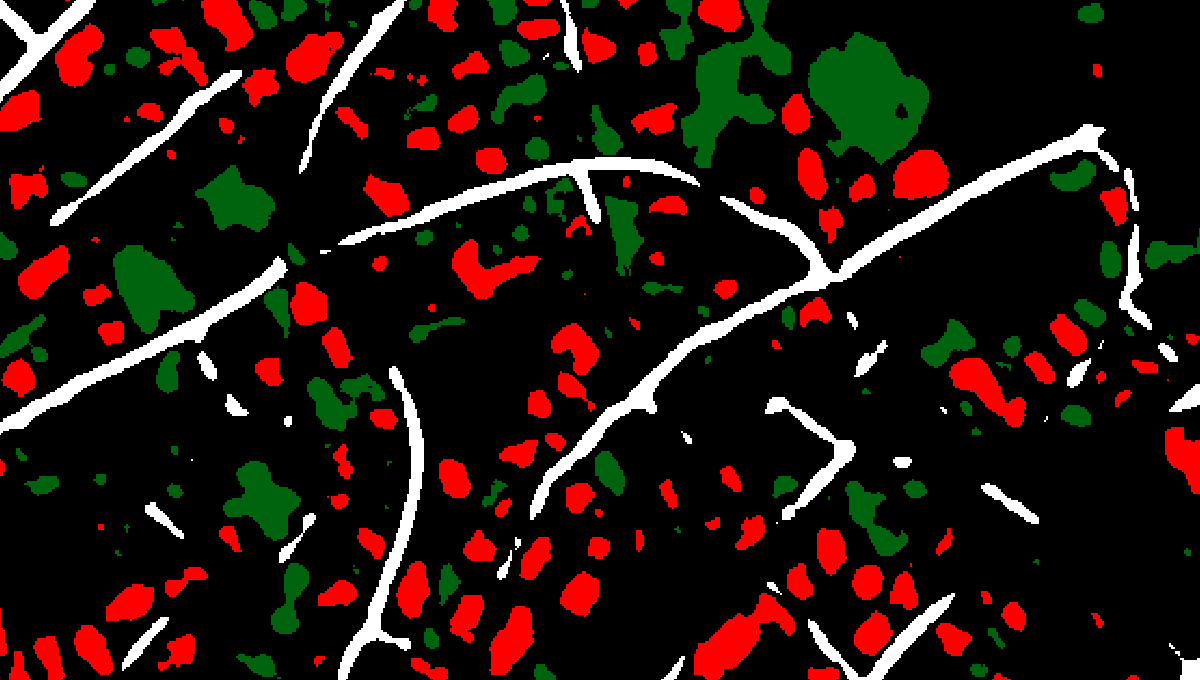}}} &
\raisebox{-.5\height}{\frame{\includegraphics[width=0.23\linewidth]{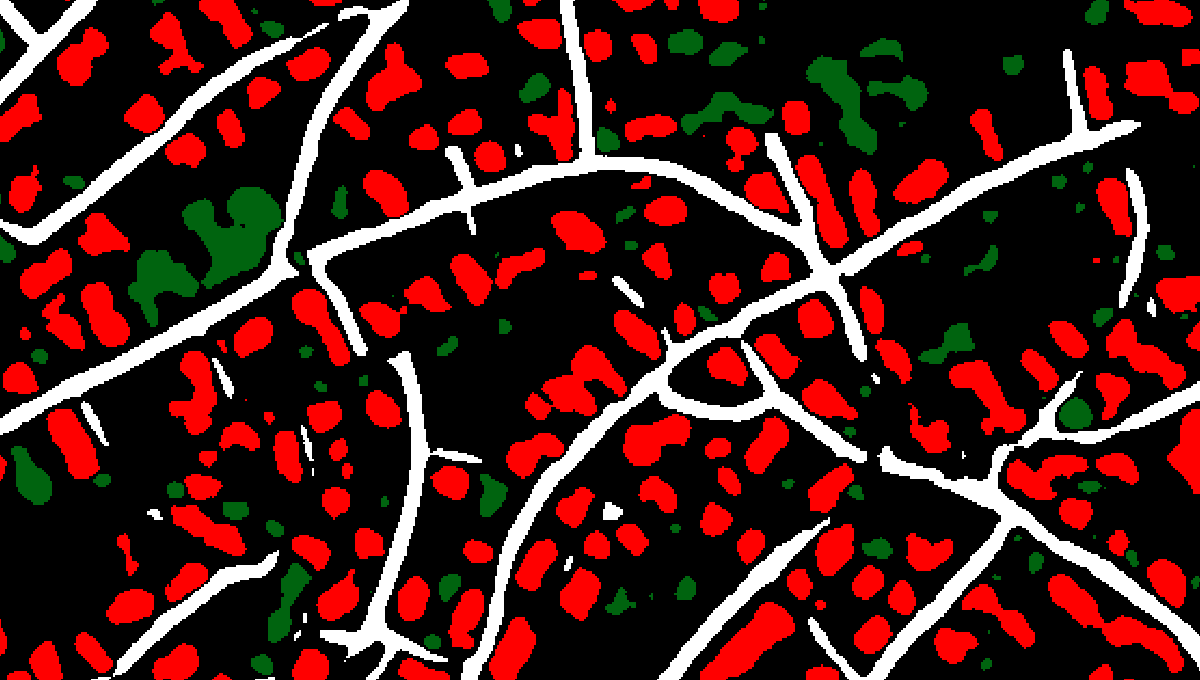}}}
\\[3.5em]
\end{tabular}
\caption{Comparison between  the traditional U-net and our framework. Red, green, and white pixels represent building, road, and tree classes, respectively. The pixels in black do not belong to any class.}
\label{fig:preds}
\end{figure*}

In the pre-processing step, we split all the cities into 256$\times$256 patches with 32 pixels of overlap. We set $\lambda_1, \lambda_2, \lambda_3$, and $\lambda_4$ in Eqs.~\ref{eq:g_loss} and \ref{eq:d_loss} to 10, 10, 1, and 1, respectively. We train StandardGAN for 20 epochs with Adam optimizer, where the initial learning rate is 0.0002, the exponential decay rates for the moment estimates are 0.5 and 0.999, respectively. In each training iteration of StandardGAN, we randomly sample 1 patch from each domain. After the 10$^{th}$ epoch, we progressively reduce the learning rate in each epoch as:
\begin{equation}\label{eq:lr}
\text{learn. rate} = \text{init\_lr} \times \frac{\text{num\_epochs} - \text{epoch\_no}}{\text{num\_epochs} - \text{decay\_epoch}}, 
\end{equation}
where init\_lr, num\_epochs, epoch\_no, and decay\_epoch correspond to the initial learning rate (0.0002 in our case), the total number of epochs (we set it to 20), the current epoch no, and the epoch no in which we start reducing the learning rate (we determine it as 10). Table~\ref{table:training_time} reports the total number of training patches in both experiments and the training time of StandardGAN. We first standardize all the data. We then train a model on the standardized source data and classify the standardized target data. We compare our approach with the other standardization algorithms described in Sec.~\ref{sec:related_work}, namely gray-world~\cite{buchsbaum1980spatial}, histogram equalization~\cite{Gonzalez}, and Z-score normalization~(Eq.~\ref{eq:z_norm}). We use U-net~\cite{ronneberger2015u} as the classifier. We also provide the experimental results for naive U-net without applying any domain adaptation methods. For each comparison, we train a U-net for 35 epochs via Adam optimizer with the learning rate of 0.0001 and the exponential decays rates of 0.9 and 0.999. In each training iteration of U-net, we use a mini-batch of 32 randomly sampled patches. We perform online data augmentation with random rotations and flips.

In Fig.~\ref{fig:fake_exp1}, we depict close-ups from the cities used in the first experiment and the fake data generated by StandardGAN. Note that to train a model, we do not use the target stylized source data, we use only the standardized data that are highlighted by red bounding boxes in the figure. The style transfer between each domain is the prior step to the standardization. We can clearly observe that there exists a substantial difference between the data distributions of the real data, whereas the standardized data look similar. Moreover, Fig.~\ref{fig:histograms} verifies that color histograms of the standardized data are considerably closer to each other than those of the real data. Fig.~\ref{fig:fake_exp6} shows closeups from the cities in the second experiment and their standardized versions by StandardGAN. The standardized and the real data for Salzburg Stadt and Lille seem quite similar. The reason is the data distributions of these two cities are already somewhere between the distributions of all five cities. However, the radiometry of Villach, Bourges, and Vaduz significantly changes after the standardization process. Besides, all the standardized data have similar data distributions.

Tables~\ref{table:ious_exp1} and \ref{table:ious_exp2} report the intersection over union (IoU)~\cite{csurka2013good} values for both experiments. The training data acquired over a single country are usually more representative for a city from the same country than a city from another country. For this reason, the quantitative results for the first experiment are generally higher. Besides, in some cases, the representativeness of the samples belonging to different classes may vary. For instance, in the first experiment, the traditional U-net already exhibits a relatively good performance for tree class, as the tree samples from the source domains represent well the samples in the target data. For this class, the performance of our method is slightly worse. It is probably because of some artifacts generated by the proposed GAN architecture when standardizing the domains. On the other hand, for the other classes, our approach achieves a better performance than all the other methods. In the second experiment, unlike the first one, none of the class samples in the source domains are representative for the target domain. Hence, the performance of U-net is poor. In addition, the common heuristic based pre-processing methods do not help improving the results. However, the StandardGAN better allow the classifier to generalize completely different geographic locations. Fig.~\ref{fig:preds} illustrates the improvement of our framework against the naive U-net in terms of predicted maps. 

\section{Concluding Remarks}

In this study, we presented novel StandardGAN, which is a new pre-processing approach proposed with the purpose of standardizing multiple domains. In our experiments, we verified that the standardized data generated by StandardGAN enable the classifier to significantly better generalize to new Pl{\'e}iades data. Note that StandardGAN has only one encoder, one discriminator, one decoder, and $n$ style encoders. Although there are multiple style encoders, their architecture is fairly simple. Thus, it is feasible to use StandardGAN to standardize larger number of domains than the number of cities in our experiments. As future work, we plan to use StandardGAN for adaptation of more domains and for other types of remote sensing data such as Sentinel, aerial, and hyper-spectral images. In addition, we plan to investigate whether StandardGAN could be used for other real-world applications such as change detection.

{\small
\bibliographystyle{ieee_fullname}
\bibliography{refs}
}

\end{document}